# A Perspective on Deep Imaging

*Ge Wang, PhD, IEEE Fellow*

*Abstract* — **The combination of tomographic imaging and deep learning or machine learning in general promises to empower not only image analysis but also image reconstruction. The latter aspect is considered in this perspective article with an emphasis on medical imaging to develop a new generation of image reconstruction theories and techniques. This direction might lead to intelligent utilization of domain knowledge from big data, innovative approaches for image reconstruction, and superior performance in clinical and preclinical applications. To realize the full impact of machine learning on medical imaging, major challenges must be addressed.**

## I. INTRODUCTION

In May 2016, *IEEE Transactions on Medical Imaging* published a special issue on "*Deep Learning in Medical Imaging*" [1]. There were 18 papers in that issue, selected from 50 submissions, showing an initial impact of deep learning on the medical imaging field. Deep learning is one of the ten breakthrough technologies of 2013 [2], and has enjoyed an explosive growth over past years; see Figure 1. As an image reconstruction researcher, I think that the special issue [1] is only the tip of the iceberg, and the potential impact of machine learning should be huge on the imaging field at large, including medical and biological imaging, industrial non-destructive evaluation, homeland security screening, and so on. In this perspective, we primarily focus on medical imaging.

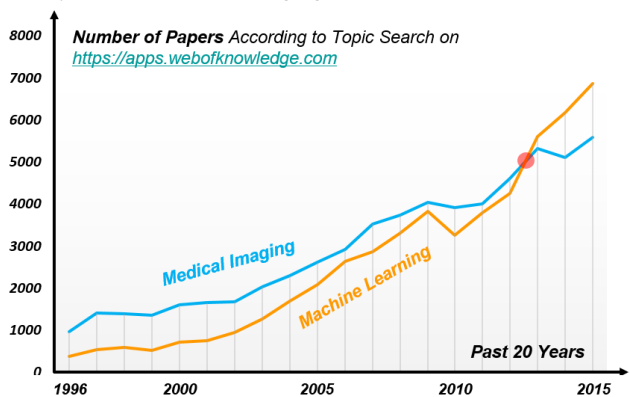

Figure 1. Attention to machine learning has exponentially increased over the past two decades, having a stronger trend than that of medical imaging. The intersection point (in red) indicates that research efforts in medical imaging and deep learning recently became comparable (fifty-fifty). Hopefully, their combination might boost both.

It is well known that there are two major components of medical imaging: (1) image formation/reconstruction: from data to images, and (2) image processing/analysis: from images to images (denoising, *etc.*) and from images to features (recognition, *etc.*). While the special issue [1] and many other publications are on image processing/analysis, there seem to be tremendous opportunities to explore the implications of machine learning for image formation/reconstruction. A big picture of the relevance of deep learning, the state of the art of machine learning, to medical imaging is given in Figure 2, which defines the concept of "*deep imaging*".

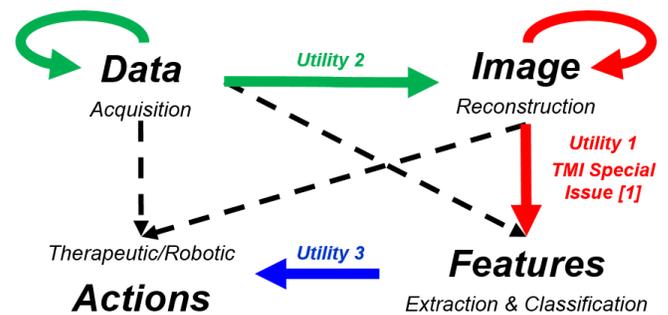

Figure 2. Big picture of deep imaging – A full fusion of medical imaging and deep learning. A high likelihood is that the direct paths from data to features and actions may need an intermediate layer essentially equivalent to a reconstructed/processed image.

Excited by the synergy between medical imaging (an exemplary aspect of tomographic imaging) and deep learning (machine learning in general), this article summarizes my opinions and speculations on deep imaging for the purposes of brainstorming, debates or initiatives in the tomographic imaging field. Although I am solely responsible for this article, many peers will be acknowledged below who stimulated my thinking and writing, and collectively we hope to set a stage for better imaging tomorrow, and medical imaging in particular.

The next section explains why the deep neural network is relevant to image reconstruction. In the third section, how to make efforts for low-hanging and high-hanging fruits is suggested with a number of key problems elaborated, and it is hypothesized that a unified deep imaging framework could be established. In the fourth section, pilot results are touched upon, showing a promise of machine learning, especially deep learning. In the fifth section, theoretical issues are discussed, including some heuristics to appreciate the potential of deep networks. In the sixth section, deep imaging as a paradigm shift is highlighted, and a conclusion is drawn.

## II. RATIONAL FOR DEEP LEARNING BASED RECONSTRUCTION

As the center of the nervous system, the human brain contains billions of neurons [3]. Neuroscience views the brain as a biological "computer" with a complicated biological neural network [3] responsible for human intelligence. In the view of an engineer, the neuron is an electrical signal processing unit. Once a neuron is excited, voltages are maintained across membranes by ion pumps to generate ion concentration differences through ion channels in the membrane. If the voltage is sufficiently changed, an action potential is triggered to travel along the axon via a synaptic connection to another neuron. The dynamics of the whole neural network is far from being fully understood. Inspired by the biological neural network, artificial neurons were designed as elements of an artificial neural network (ANN) [4]. This model linearly combines data at input ports like dendrites, and non-linearly transforms the weighted sum into the output port like the axon.

While the ANN approach was well motivated, for about two decades this and other machine learning techniques had not caused the public excitement until deep learning became the buzzword years ago. One of the criticisms of neural networks had been the need for extensive data — the training time that scales poorly with network size and problem complexity, and the risk that model identification could be trapped at a local extremum. Last year in a presentation given at Cambridge University, Dr. Geoffrey Hinton of University of Toronto explained how the deep neural network made an exciting breakthrough. Briefly speaking, the enabling factors are multiple: thousands of times more data (big data), millions of times faster computing power, smarter weight initialization, better non-linear transformation, and significantly deeper network topology. As a remarkable milestone [5], an unsupervised learning procedure for a restricted Boltzmann machine (RBM) can be efficiently and recursively used to prepare a deep network layer by layer without supervision. Then, the pre-trained parameters can be fine-tuned via backpropagation. The successes of deep networks are now well reported in the areas of computer vision, speech recognition, language processing, and classic and electronic gaming, with the recent highlight "*AlphaGo*" (the computer program that plays the board chess Go and defeated the professional player for the first time) [6]. There are several excellent review articles on deep learning. Three complementary examples are [7], [8] and [9] (the last one is the most comprehensive and up-to-date textbook). Also, the 2015 Medical Image Computing and Computer Assisted Intervention Society (MICCAI) Conference had excellent tutorials on deep learning and imaging [10]. Since this field is rapidly expanding, an exclusive review is beyond the scope of this article.

Instead of covering much technical details of machine learning, let us first look at a neural network for pattern recognition tasks such as face recognition. As shown in Figure 3, there are many layers of neurons with inter-layer connections in the deep network. Data are fed into the input layer of the network. Weights associated with the neurons are typically obtained in a pre-training and fine-tuning process or a hybrid training process with a large set of unlabeled and labeled images. Results are obtained in the output layer of the network. Other layers are hidden from direct access. Each layer uses features from the previous one to form more advanced features. At earlier layers, more local features such as edges, corners, and facial motifs are analyzed. At later layers, more global features are synthesized to match face templates. Thanks to innovative algorithmic ingredients that have been developed over the past years, this deep learning mechanism has been made extremely successful for feature extraction from images as reported in the literature [7], [8] and [9]. Note that a deep network is fundamentally different from many other multi-resolution analysis schemes and optimization methods. A niche of deep networks is the nonlinear learning and nonconvex optimization ability for problems of huge dimensionality that were too complex for machine intelligence.

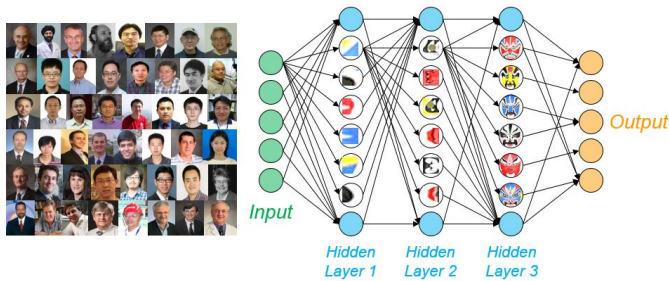

*Figure 3. Deep network for feature extraction and classification through nonlinear multi-resolution analysis.*

While Figure 3 illustrates the process *from images to features*, what we are now interested in is from projection data/indirect measurements to tomographic images. The raw data collected for tomographic reconstruction can be considered as image features which are often approximated as linearly combined image voxel values, and more generally modeled as nonlinear functions of an image. Thus, image reconstruction is *from raw data (features measured with tomographic scanning) to underlying images*, an inverse of the pattern recognition workflow that goes from images to features in Figure 3. It is argued as follows that this inverse process does not present any conceptual challenge.

A classic mathematical foundation of artificial neural networks is the so-called universal approximation theorem that, with a reasonable activation function, a feed-forward network containing only a single hidden layer may closely approximate an arbitrary continuous function on a compact subset when the network parameters are optimally specified [11]. Then, the assumption on the activation function was greatly relaxed, leading to the observation that "*it is not the specific choice of the activation function, but rather the multilayer feedforward architecture itself which gives neural networks the potential of being universal learning machines*" [12]. Although a single hidden layer neural network can approximate any function, it is highly inefficient to handle large-scale problems and big data since the number of neurons would grow exponentially. With deep neural networks, depth and width can be combined to efficiently represent functions to high precision, and perform powerful multi-scale analysis, quite like wavelet analysis but in a nonlinear manner.

If we consider that the process from images to features is a forward function, the counterpart from features to images is an inverse function. It might appear that the analogy is somehow asymmetric, since the features of the neural network are based on the problem and hence semantic, while the tomographic data acquisition captures physical interactions. However, at a higher level the information flows in semantics and physics are quite the same and should be computable in similar steps. Just like the case in which such a forward function has been successfully implemented in a neural network for many applications, so should be the case that the inverse function for various tomographic modalities would be representable in terms of a neural network. Both types of the processes ought to be feasible by the intrinsic capabilities of the deep network that supports such a general functional representation via biomimicry, be it forward or inverse. Since the forward neural network is deep (i.e., many layers from an image to features), it is natural to expect that the inverse neural network should be also deep (many layers from raw data to an image). Despite special cases in which relatively shallow networks may work well, the neural network would be generally deep when the problem is complex, and the aforementioned representation efficiency and multi-resolution analysis is important to combat the entanglement of features and the curse of dimensionality.

This viewpoint can also be argued from an algorithmic perspective. Either the filtered backprojection (FBP) or simultaneous algebraic reconstruction technique (SART) can be easily formulated in the form of parallel layered structures [13]. Then, the straightforward path to deep imaging could be simply from raw data to an initial image through a neural network modeled after a traditional reconstruction scheme, and then from a reconstructed image to a processed image through a refinement deep network (an overlap with deep learning based image processing, to be explained more in the next section).

As a side note, from data sampled below the traditional Nyquist rate using the Fourier/Radon methods, artifacts in reconstructed images are frequently structured and non-local. It might appear that a deep convolutional neural network is better at handling localized artifacts than it is in the case of image distortions with a non-local spatially-varying point spread function. Actually, nonlocal features can be expressed as a linear or nonlinear combination of local features. Hence, deep learning can handle global image features in principle, even if they are substantially distorted.

## III. ROADMAP FOR LOW- AND HIGH-HANGING FRUITS

Without the loss of generality, let us take CT as an example. It can be imagined that many CT reconstruction algorithms can be covered in the deep imaging framework, as suggested in Figure 4. In the past, image reconstruction is mostly analytic, and advanced reconstruction algorithms were developed even in the intricate helical cone-beam geometry assuming noise-free data [14]. With the increasing use of CT scans and public concern about patient radiation safety, iterative reconstruction algorithms have become gradually popular [15]. It is hypothesized here that both analytic and iterative algorithms can be upgraded to deep imaging algorithms to deliver superior diagnostic performance.

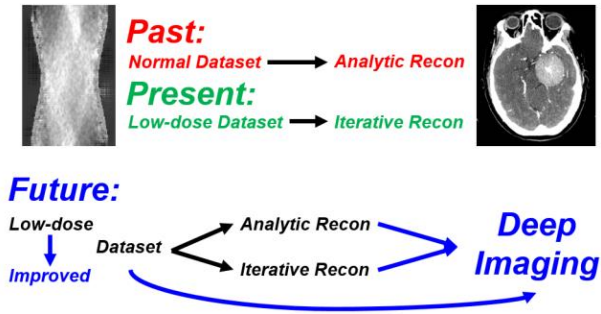

*Figure 4. Past, present and future of CT image reconstruction.*

Towards deep imaging, there are various techniques to be developed as low-hanging and high-hanging fruits. Let us first look at low-hanging fruits, and then high-hanging ones. However, these can be pursued in parallel.

### III.1. Low-hanging Fruits

First, low-hanging fruits can be harvested by replacing one or more machine learning elements of a current image reconstruction scheme with deep learning counterparts. To appreciate this replacement strategy, let us recall genetic engineering techniques. Geneticists use knock-out, knock-down and knock-in to produce genetically modified models such as genetically modified mice. In a nutshell, knock-out means deletion or mutational inactivation of a target gene; knock-down suppresses the expression of a target gene; and knock-in inserts a gene into a chromosomal locus. Once a target gene is "knocked-out", it no longer functions. By identifying the resultant phenotypes, the function of that gene can be inferred. Less brutal than knock-out, knock-down weakens the expression of a gene. On the other hand, knock-in is just the opposite of knock-out. In a similar spirit, we can view each type of reconstruction algorithms as an organic flowchart, and some building blocks can be replaced by machine learning. For example, Figure 5 shows a generic flowchart for iterative reconstruction, along with multiple machine learning elements that can be knocked-in at appropriate locations while the corresponding original black box can be knocked-out and knocked-down. Thus, a state of the art reconstruction algorithm can be used to guide the construction of a corresponding deep network. By the universal approximation theorem, each computational element should have a neural network counterpart. Therefore, a network-oriented equivalent version can be built out of the state of the art algorithm. The real power of the deep learning based reconstruction lies in the data-driven knowledge-enhancing abilities so as to promise a smarter initial guess, more relevant intermediate features, and an optimally regularized final image within an application-specific low-dimensional manifold.

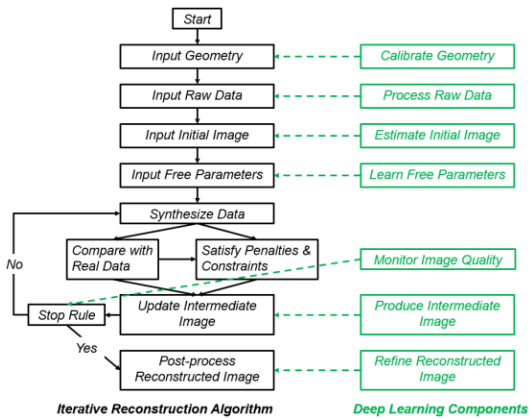

*Figure 5. Low-hanging fruits by "knocking-out/down/in" computational elements in a traditional iterative reconstruction flowchart.*

Also, low-hanging fruits can be captured by performing deep learning based image post-processing (a number of good papers were already published on image denoising using neural networks; however, few of which are for medical imaging [1]). When a projection dataset is complete, an analytic reconstruction would bring basically a full information content from the projection domain to the image space even if data are noisy. If a dataset is truncated, distorted or otherwise severely compromised (for example, limited angle, few-view, local reconstruction, metal artifact reduction, beam-hardening correction, scatter suppression, and motion restoration problems) [16], a suitable iterative algorithm can be used to form an initial image. It is the image domain where the human vision system is good at denoising, destreaking, deblurring, and interpretation. In other words, we can let existing image reconstruction algorithms generate initial images, and then let deep networks do more intelligent work based on initial images. This two-stage approach is recommended for three reasons. First, well-established tomographic algorithms can still be utilized. Second, the popular deep networks with images as inputs can be easily adapted. Third, domain-specific big data can be incorporated as unprecedented prior knowledge. With this approach, the neural network is naturally deep, since as shown in many papers [1] medical image processing and analysis can be effectively performed by a deep network.

Similarly, a sinogram can be viewed as an image, and a deep learning algorithm can be used to improve a low-dose or otherwise compromised sinogram; see an example we simulated below. The transform from a poor sinogram to an improved one is a type of image processing tasks, and can be done via deep learning. Then, a better image will be reconstructed from the improved sinogram.

### III.2. High-hanging Fruits

In contrast, high-hanging fruits do not necessarily involve any key ingredient of classic reconstruction algorithms. With the most advanced deep imaging algorithms, we hope to encompass the broadest range of image reconstruction problems for an imaging performance superior to the state of the art. It seems that the following key problems are of great reference values (some of which you might still consider as low-hanging fruits), and solving them with novelty, thoroughness, and applicability would lead to high-hanging fruits.

<u>Network Configuration</u> – Design of network topologies (and dynamics) for typical applications is a prominent target, along the formulation of the working principles (like object-oriented design) for adaptation and integration of network modules. This is equivalent to algorithmic design or computer architecture design. It may be boldly speculated that deep imaging networks could potentially outperform conventional imaging algorithms, because information processing with a deep network is nonlinear in activation functions, global through a deeply layered structure, and a best bet with comprehensive prior knowledge learnt from big data. This is in sharp contrast to many traditional regularizers that are linear, local, or *ad hoc* [17]. Currently, the network design remains an area of active exploration in terms of both the overall architecture and component characteristics, and has been rarely touched upon for the purpose of image reconstruction.

The deep neural network, and artificial intelligence in general, can be further improved by mimicking neuroplasticity — the ability of the brain to grow and reorganize for learning, adaption and compensation. Currently, the number of layers and the number of neurons per layer in a deep network were obtained using the trial and error approach, and not governed by any theory. In reference to the brain growth and reorganization, the future deep network could work in the same way and become more adaptive and more suitable for medical imaging. As time goes by, we may be able to design deep networks that are time-varying, reconfigurable, and even have quantum computing behaviors [18].

<u>Data Generation</u> – Advanced image modeling and data generation are important. In the clinical world, there are enormous image volumes but only a limited amount of them were labeled, and patient privacy has been a hurdle for medical imaging research. Nevertheless, the field is ripe for big data and deep learning. First, big data are gradually becoming accessible to researchers. A good example is the National Lung Screening Trial (NLST) [19]. In this context, pairing imaging data with reconstructed images is invaluable. On the other hand, a really realistic simulator could play a key role as well. For example, a high-performance simulator (such as CatSim for CT research [20]) can take real images as input to produce high-quality "raw data" and "labeled" images for training and testing purposes. More interestingly, we can build a general anatomical image model to generate big data. For example, using anatomical atlases such as those based on the Visible Human project [21, 22], we can produce image volumes that are representative of the human bodies in different contrasts (say CT and MRI). With deformable morphing methods, we can produce a very large number of anatomically realistic image data that may be otherwise difficult to obtain [23]; see Figure 6.

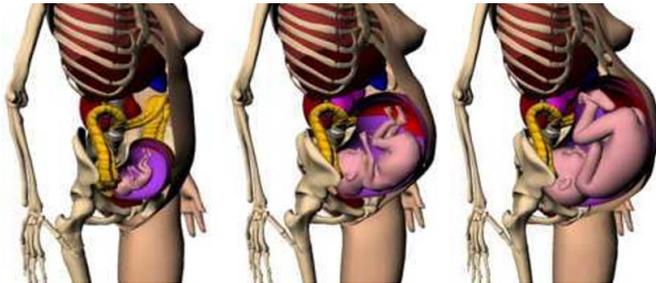

*Figure 6. Organ and body surfaces of the RPI Pregnant Female models at the end of 3, 6 and 9 months, respectively, for the purposes of estimating radiation doses from radiological procedures [23] (Courtesy of X. George Xu with RPI).*

It is underlined that simple-minded synthetic big data, based on one or a few real cases only, would be insufficient for deep imaging. Indeed, the current standard for high-quality image reconstruction papers is to have real data demonstration, not just numerical results. Existing image databases do not store raw data, and preprocessing steps and parameters, and a typical numerical simulation environment in the medical imaging field does not include all practical factors including x-ray scattering, x-ray focal spot shape, detector or coil imperfectness, physiological motion, and many others. However, when there are sufficiently many representative cases that allow us to deform into realistic variants, we can generate a huge number of unlabeled and labeled high-quality images as target. Coupled with low-quality/incomplete data/images as input, we can train a deep neural network to perform image reconstruction that could bypass an explicit treatment of incomplete imaging physics details. This is quite like the case in which a deep learning based diagnosis program designed by a computer scientist can be comparable or even better than a human radiologist, although the computer scientist knows little about pathology [1].

<u>Hybrid Training</u> – Of particular relevance to deep imaging is how to train a deep network with big data. With unlabeled big data and a smaller or moderate amount of labeled data, deep learning can be implemented via a pre-training step without supervision, a knowledge transfer based initialization, or a hybrid training process, so that intrinsic image features are learnt to have favorable initial weights that are then fine-tuned [24]. Transfer learning and hybrid training with unlabeled and labeled data seems a good research topic. For example, such a training process could be pre-conditioned or guided by an advanced numerical simulator, observer, and statistical bootstrapping.

<u>Network-based Regularization</u> – Smart regularization in the data and/or image domains can be viewed as a generalized "backpropagation". Some network modules are needed to extract desirable/undesirable image features and enable the network to do smart image reconstruction. In particular, some modules are needed to effectively regularize image reconstruction to reduce task-specific penalty measures. Evaluation of penalties is relative easier than generalized backpropagation of these penalty measures, both of which are desirable to improve image quality gradually.

"<u>*Particulars*</u>" Management – In many imaging modalities, data quality and image metrics are complicated by multiple factors, such as imaging geometry, patient placement, sensor calibration, and so on. Prior information about these characteristics was taken into account, to different degrees, in existing reconstruction approaches but it is not straightforward within the deep learning framework. Recently, with locally linear embedding we obtained excellent results in automatic geometric calibration for fan-beam and cone-beam CT [25]. Based this and other results, preprocessing/calibration tasks could be handled in a low-dimensional manifold, and seem computationally manageable in a deep learning framework.

The above considerations apply to all major medical imaging modalities, since these biomedical imaging problems are associated with similar formulations in the general category of inverse problems. To the first order approximation, a majority of medical imaging algorithms have Fourier or wavelet transform related versions, and could be helped by some common deep networks. For nonlinear imaging models, deep imaging could be a better strategy, given the nonlinear nature of deep networks. While the multimodality imaging trend promotes a system-level integration [26], deep imaging might be a unified information theoretic framework or a meta-solution to support either individual or hybrid scanners.

The suggested imaging algorithmic unification is consistent with the successes in the artificial intelligence field where deep learning procedures follow very similar steps despite the rather different problems such as chess playing, face identification, and speech recognition. Just as a unified theory is preferred in the physical sciences, it is speculated that a unified medical imaging methodology would have advantages so that important computational elements for network training and other tasks can be shared by all the modalities, and the realization of inter-modality synergy could be facilitated since all the computational flows are in the same hierarchy consisting of building blocks that are standard and dedicated artificial neural circuits.

Additionally, some modalities have characteristics that are complex- or tensor-valued, while most deep learning architectures have been real-valued. There are simple ways to convert from complex to real, and there may be value in developing neural network architectures that support complex- or tensor-valued inputs and outputs. Contemporary tensor decomposition methods could be used to motivate deep network structures [27].

Certainly, not every imaging problem can be best solved using deep learning. For a clean dataset, the conventional method works well. For a challenging dataset, a deep network may be the method of choice. In any case, deep learning should be very relevant to medical imaging. Like with other methods, research will uncover the merits and limits of deep imaging, and future image reconstruction schemes may be hybrid, without discarding classic results entirely.

Along the course of deep imaging development, the first step is to show the technical feasibility that deep learning is qualified as an alternative approach; the second step is to achieve a statistically better image quality from deep learning than competing methods; and the third step is to make deep learning solutions highly efficient and practical for deployment.

## IV. PILOT RESULTS

There have been a number of initial attempts at using machine learning and deep learning for medical image reconstruction. With the use of neural network, two classical papers from more than 20 years ago targeted SPECT image reconstruction [28, 29]. More recently, dictionary learning, which is a contemporary machine learning approach, was adapted for MRI and CT image reconstruction [27, 30, 31]. Very recently, reports came out on pilot MRI results with initial images and regularization parameters respectively obtained via deep learning [32, 33], and on limited-angle CT via data-driven learning based on a deep neural network, showing artifacts reduction and detail recovery [34].

Our three CT examples are as follows. The first example shows how to transform a poor-quality image to a good-quality counterpart. Let us define a 2D world of Shepp-Logan phantoms. Let a field of view be a unit disk covered by a 128*128 image, 8 bits per pixel. We made each image consist of one background disk of radius 1 and intensity 100 as well as up to 9 ellipses completely inside the background disk. Each ellipse is specified by the following random parameters: center at (x, y), semi-axes (a, b), rotation angle θ, and intensity selected from [-10, 10]. A pixel in the image could be covered by multiple ellipses including the background disk. The pixel value is the sum of all the involved intensity values. From each image generated, 256 parallel-beam projections were synthesized, 180 rays per projection. From each dataset of projections, a SART reconstruction was performed for a small number of iterations. These blurry intermediate images are not what we want. Then, a deep network was trained using the known original phantoms to predict a much-improved image from a low-quality image. The network consisted of three convolutional layers. In the training process, 140,000 small image patches were randomly selected from the intermediate images as input, and the corresponding image patches in the ground truth images as output. The representative results are in Figure 7.

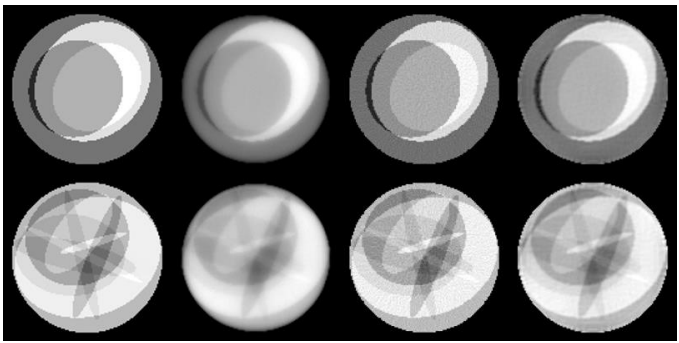

*Figure 7. Deep network capable of iterative reconstruction. The image pairs from the left to right columns are (1) two original phantoms, (2) the SART reconstruction after only 20 iterations, (3) the counterparts after 500 iterations, and (4) the deep imaging results with the corresponding 20-iteration images as the inputs, which resemble well the 500-iteration counterparts (arguably, the 4th column looks slightly better than the 3rd column).*

The second example is from a poor-quality sinogram to a good-quality sinogram, which was prepared in a way quite similar to that for the first example. Now, each phantom contained a fixed background disk and two random disks inside the circular background: one disk represents an x-ray attenuating feature, and the other an x-ray opaque metal part. The image size was 32x32 for quick results. After a phantom image was created, the sinogram was generated from 90 angles. Every metal blocked sinogram was linked to a complete sinogram formed after the metal was replaced with an x-ray transparent object. Then, a deep network was similarly trained as for the first example with respect to the complete sinograms to restore missing data; as illustrated in Figure 8.

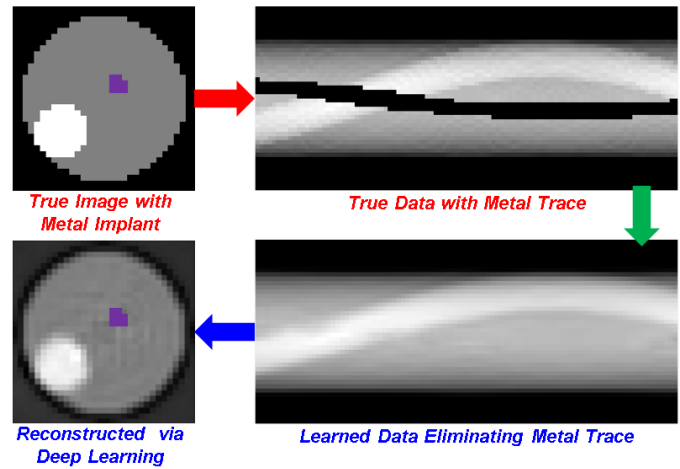

*Figure 8. Deep network capable of sinogram restoration. The first row lists an original image (metal in purple) and the associated metal-blocked sinogram. The second row contains the restored sinogram and the reconstructed image, which shows the potential of deep imaging as a smart interpolator over the missing data region.*

The third example demonstrates the potential of deep learning with MGH Radiology chest CT datasets [21]. These datasets were acquired in low dose levels. They were reconstructed using three reconstruction techniques: filtered back-projection (FBP), adaptive statistical iterative reconstruction (ASIR), and model-based iterative reconstruction (MBIR) respectively. These were all implemented on commercial CT scanners. We followed the same deep learning procedure as in the previous two examples, and took FBP images as the input and MBIR images as the gold standard for neural network training. For comparison, we performed image denoising on FBP images using the famous block matching and 3D filtering (BM3D) method [35] and our deep neural network. Figure 9 shows the image denoising effect of deep learning, as compared to the MBIR counterpart. It can be observed that the image quality achieved via deep learning is quite similar to that of MBIR but the result from deep learning is much faster than the state of the art iterative reconstruction. It is interesting that a computationally efficient post-processing neural network after the standard "cheap" FBP achieves the same outcome as the much more elaborative iterative scheme, and yet the neural network solution does not need any explicit physical knowledge such as the x-ray imaging model. We are working to refine the network so that deep learning might beat the iterative reconstruction, aided by richer knowledge extracted from big data.

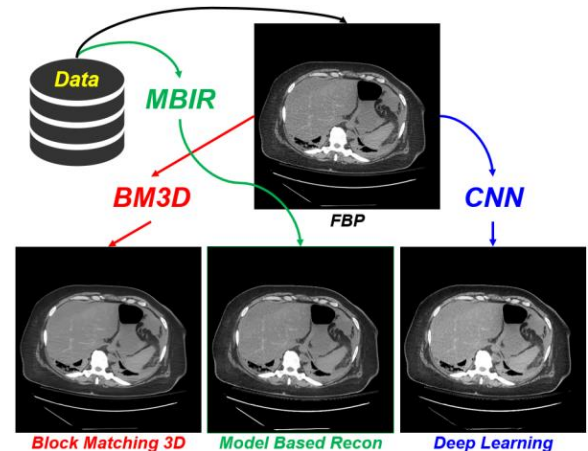

*Figure 9. Deep learning based image denoising, demonstrating that deep learning could be an effective and efficient alternative to the state of the art iterative reconstruction strategy.*

Figure 7 resembles a super-resolution problem, Figure 8 represents a type of imprinting work. Figure 9 is just an image denoising operation. Note that you may not be able to see much differences between the iterative and deep learning images in Figure 9 partially because the iterative results were used as the target for training, but the point is that if deep imaging and the iterative method give similar results, then deep imaging is already an alternative method, and more to expect in the future. Indeed, it is relatively easy to apply existing deep learning techniques to have these low-hanging fruits first. Indeed, a latest report of ours visually and quantitatively demonstrated a competing performance of deep learning relative to TV minimization, KSVD, and BM3D [36].

In principle, such artificial network based *a priori* knowledge is more powerful than convex approaches, as hinted by the superior performance of deep learning in well-known image analysis applications such as face recognition. There are multiple ways to capitalize the power of deep learning for image reconstruction. The use of neural networks as "priors" in an algebraic formulation is one of them. We have already pursued along that direction; for example, with a neural network based stopping rule the SART-type reconstruction can be "smartly" monitored (data not included due to limited space).

It is encouraging that in addition to what was mentioned above there are other researchers who are working on deep learning approaches (including the author's collaborators and peers). However, up to this moment, no deep learning based reconstruction method has been fully developed, rigorously validated, and has outperformed the best alternative methods. The challenges are very understandable. First of all, deep learning based imaging needs high-quality big data and meticulous training and testing for full characterization of the imaging performance. This motivates us to take solid steps immediately for low- and high-hanging fruits.

## V. THEORETICAL ISSUES

Deep learning has achieved impressive successes in practice but a decent theory remains missing. This embarrassment is well recognized by the community. For example, Dr. Yann LeCun pointed out several open topics in his keynote presentation at the *Conference on Computer Vision and Pattern Recognition 2015*. Several of them demand deeper theoretical understanding on why CNN works well and how to make it better so that local minima are effectively avoided and global solutions are efficiently achieved.

Why does a deep neural network perform well? This has become a hot topic for theorists to brainstorm. For example, from a perspective of theoretical physics, the concept of the renormalization group (RG, related to conformal invariance by which a system behaves in the same way at different scales) was recently utilized for understanding the performance of deep learning [37]. Specifically, a mapping was constructed in light of the variational renormalization group for the deep learning network based on Restricted Boltzmann Machines (RBMs), suggesting that deep learning may be an RG-like scheme to learn features from data.

Here let me give an insight from a perspective of linear/non-linear systems. First, each neuron is governed by an activation function which takes data in the form of an inner product, instead of input data directly. The inner product is computed by summing products of paired data, which can be visualized as a double helical structure, like that of DNA in which the paired results between the double helix are lumped together. In other words, the inner product is a fundamental construct for deep learning. This view is mathematically meaningful, since most of mathematical transforms including matrix multiplications are calculated via inner products. These products are nothing but projections onto appropriate bases of an involved space. Cross- and auto-correlations are inner products, common for feature detection. Projections and backprojections are inner products as well. Certainly, the inner product operation is linear, and we should not be limited to linearity. Then, the nonlinear trick comes as an activation function. Biomimicry-wise, the biological convolution (the function of DNA) is much more complicated than the classic mathematical convolution, and that is why the nonlinear activation function is needed to empower the classic mathematical convolution. In a good sense, a deep network is a generalized mathematical convolution process through multiple stages. Since a deep network mimics an organism better than a linear operator, a deep network is much more intelligent than a linear system solver.

In a deep network, the alternating linear and nonlinear processing steps seem to hint that the simplest linear computational elements (inner products) and simplest nonlinear computational elements (monotonic activation functions) can be integrated to perform highly complicated computational tasks. Hence, the principle of simplicity applies not only to physical sciences but also to information/intelligence sciences, and the multi-resolution phenomena seems merely a reflection of this principle. When inner products are performed, linear elements of machine intelligence are realized. When the activation steps (in a general sense, many effects are included such as pooling and dropout) are followed, the non-linear nature of the problem is addressed. So on and so forth, smart analysis and synthesis goes from bottom up (feed forward) and from top down (back propagation). *This kind of linear and nonlinear couplings/interconnections might universally solve a wide class of nonlinear optimization/estimation/intelligence problems, whose theoretical characterization has yet to be worked out.*

As a side note, a majority of reconstruction algorithms were designed for linear imaging problems. If the linear system model is accurate, at the first look, there appears no need to trade analytic and statistical insight for nonlinear processing advantages of deep networks. Nevertheless, even in that case, it could be argued that deep imaging is an attractive platform to fully utilize domain specific knowledge when big data is available. Such comprehensive contextual prior knowledge cannot be fully utilized by iterative likelihood/Bayesian algorithms, which are nonlinear but limited to compensation for statistical fluctuation. Additionally, with the principle of simplicity, we tend to prefer deep imaging, using the analogy of digital over analog computers.

It is acknowledged that interesting critiques were made on deep learning that slightly different images could be put into distinct classes [38], and random images could be accepted into a class with a high confidence level [39].

These critiques are important, but they should be addressable. Taking CT as an example, image reconstruction is theoretically not unique from a finite number of projections in an unconstrained functional space [40]. However, the non-uniqueness is avoided in practice where we have a priori knowledge that an underlying image can be treated as band-limited, and we can collect a set of sufficiently many data appropriate for the bandwidth. As another example in the area of compressed sensing, it was shown that while compressed sensing produced visually pleasing images, tumor-like features were hidden or lost [41]. Nevertheless, these features were constructed based on the known imaging geometry and algorithmic details, which would not likely be encountered in clinical settings. Indeed, most theoretical analyses on compressed sensing methods suggest the validity of the results with the modifier "*with an overwhelming probability*", such as in [42]. Actually, multiple iterative image reconstruction algorithms for medical imaging already have CS components and show excellent results. As long as a method deliver decent results most likely, it is a great tool unless we have an even better method.

Still, there are more theoretical limitations of compressed sensing that have yet to be resolved. When the claim was made that compressed sensing generates valid results "*with an overwhelming probability*", important caveats cannot be ignored. Especially, the problem sizes need to be large for most theoretical results to become valid, and the probabilistic sampling schemes have to be generated according to distributions that may not be easily achievable or verifiable. Even if there is a high-probability of "success" in the theoretical settings, the involved constants of proportionality are not always favorable. Although the current theory cannot give the imaging performance guarantee for most medical imaging problems, the theoretical insights have enabled a large range of applications.

Overall, we feel that the story for deep learning will be similar to that for compressed sensing; that is, dependably-good results are feasible in the absence of full-fledged theory, and eventually we will have a satisfactory theory. Encouragingly, good results are constantly emerging such as [43].

It is also acknowledged that deep imaging has important implications regarding how image details get resolved, and its risk to bring details or artifacts that are not purely in the data. This could be a huge concern in medicine. This precaution applies to all other regularization-based algorithms as well, although to different degrees. Traditional regularization methods were extensively studied already, and there are opportunities for deep learning research along this direction. Philosophically and practically, I think that we should be able to reconstruct an optimal image from adequate measurements in reference to rich prior knowledge using a deep network, its variants, or other similar methods. Many regularized iterative algorithms were demonstrated to be successful, and I do not fundamentally worry about that deep learning based algorithms will be cheated. A key prerequisite for deep imaging is a training set that spans the space of all relevant cases. Otherwise, an optimized deep network topology (if achieved) could be disappointing in the real world.

If a deep network is well trained, it is postulated that its structure should be stable through re-training with images obtained through locally and affinely transformed previously-used images. This invariance may help characterize the generic architecture of a deep imager.

## VI. CONCLUDING REMARKS

In 1962, Dr. Thomas Kuhn proposed a philosophical view of scientific advancement [44, 45]. Instead of viewing science as a steady progress in incremental steps, he underlined breakthroughs, each of which is characterized by a new paradigm of thinking and doing. In 2007, Dr. Jim Gray gave a talk to the *National Research Council, Computer Science and Telecommunications Board*, in Mountain View, CA, in which he added the fourth paradigm to the perspective from the pre-science era to the present: empirical, theoretical, computational, and data-explorative that unifies theory, experiment, and simulation; see Figure 10. Actually, the fourth paradigm might be better called "machine learning from big data" to emphasize that the driver of big or small data exploration is the man-machine system, instead of the researcher alone; i.e., intelligence is no longer solely owned by the human. By the way, the fifth paradigm has not been mentioned yet, which I think should be hybrid (brain-computer integration) yet connected intelligence (Intelligent-net or "*Intelnet*") enhancing learning and research capabilities to an unprecedented level.

The fourth paradigm seems to be on the horizon in the medical imaging field. In addition to exploring how deep learning can reshape the landscape of image reconstruction as pondered earlier, the combination of deep learning based image reconstruction and analysis may allow us to change healthcare. For example, deep learning may help tweak or design imaging and reading protocols specific to different organs, lesion types, and patient characteristics. Also, big data based deep imaging software may query data across institutional and medical specialties and go beyond the existing decision support programs by incorporating such information as patient age, gender, symptoms, medical history, disease profile, biochemical, pathological, microbiological, and genomic data. Moreover, why not combine diagnosis and intervention via deep learning? Supervised autonomous robotic soft tissue surgery is an initial example [47]. There are already efforts to automate radiation treatment planning. This kind of intelligent systems will be a counterpart of the *GoogleCar* (which is the automatic driving car being developed by Google): the former inside, the latter outside.

Deep learning is not only a new wave of research, development and application in the field of medical imaging (and other imaging fields such as homeland security screening) but also a paradigm shift. This could be the magic wand to achieving optimal results cost-effectively, especially for huge and compromised data, as well as for problems that are nonlinear, nonconvex, and overly complex. However, my perspective of deep imaging could be overly optimistic, and must be balanced by controversies, potential difficulties and justified concerns. It has taken decades for the neural network approach to outperform the human in some recognition tasks, and hence the success of deep learning for image reconstruction might need some new twists that take time to develop and realize. The big data suitable for reconstruction, learning architecture, performance evaluation, and potential translation may together demand significant efforts. Nevertheless, I am enthusiastic that the venture to deep imaging will accelerate to a level that it will re-invent the future of healthcare [48, 49].

### ACKNOWLEDGMENT

The author is grateful for inspiring discussions with Drs. James Brink (Harvard MGH), Wenxiang Cong (RPI), Juergen Hahn (RPI), Michael Insana (UIUC), Nadeem Ishaque (GE GRC), Mannudeep K. Kalra (Harvard MGH), Xuanqin Mou (Xi'an Jiaotong Univ., China), Jiantao Pu (Univ. of Pittsburg), Dinggang Shen (Univ. of North Carolina), Michael Vannier (Univ. of Chicago), Yan Xi (RPI), George Xu (RPI), Pingkun Yan (Philips), Hengyong Yu (Univ. of Massachusetts Lowell), Junping Zhang (Fudan Univ., China), Yi Zhang (Sichuan University), and other colleagues who made valuable comments. The author also acknowledges the advice from an anonymous reviewer who gave advice on the theoretical limitations of compressed sensing. Figures 7, 8 and 9 were produced by Mr. Qingsong Yang (RPI).

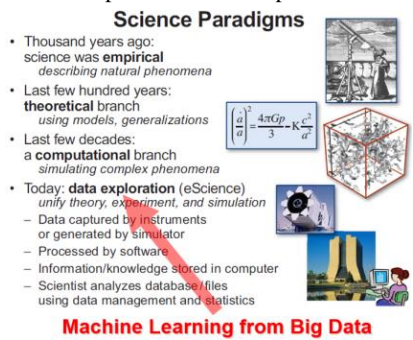

*Figure 10. Fourth paradigm as presented by Dr. Jim Gray [46], and slightly renamed by the author.*

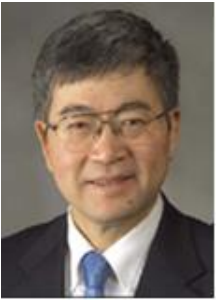

Ge Wang (F'03), PhD in ECE, is the Clark & Crossan Endowed Chair Professor and Director of Biomedical Imaging Center, Rensselaer Polytechnic Institute, Troy, New York, USA. His interest includes x-ray computed tomography (CT), optical molecular tomography, multi-modality imaging, and deep learning. He wrote the pioneering papers on the first spiral/helical cone-beam CT reconstruction algorithm (1991). Spiral/helical cone-beam/multi-slice CT is constantly used in almost all hospitals worldwide. There are >100 million CT scans yearly with a majority in the spiral/helical cone-beam/multi-slice mode. He and his collaborators published the first paper on bioluminescence tomography, creating a new area of optical molecular tomography. His group published the first papers on interior tomography and omni-tomography for grand fusion of all relevant tomographic modalities ("*all-in-one*") to acquire different datasets simultaneously ("*all-at-once*") with simultaneous CT-MRI as an example. His results were featured in *Nature*, *Science*, and *PNAS*, and recognized with academic awards. He published over 400 journal papers, some of which are highly cited (http://scholar.google.com/citations?user=pjK2mQwAAAAJ). His team has been in collaboration with world-class groups, and continuously well-funded by federal agents. He is the lead guest editor of four *IEEE Trans. Medical Imaging* special issues on cone-beam CT, molecular imaging, compressive sensing, and spectral CT respectively, the founding Editor-in-Chief of International Journal of Biomedical Imaging, Associate Editor of *IEEE Trans. Medical Imaging*, *IEEE Access*, *Medical Physics*, and *Journal of X-ray Science and Technology*. He is Fellow of *IEEE*, *SPIE*, *OSA*, *AIMBE*, *AAPM*, and *AAAS* (ge-wang@ieee.org).